\documentclass[aps,superscriptaddress,showpacs,floatfix]{revtex4}

\usepackage{amsmath}
\usepackage{amssymb}
\usepackage{bm}
\usepackage{graphicx}

\begin{document}

\title{Local level statistics for optical and transport properties of disordered                                   
systems at finite temperature
}

\author{A. Malyshev
\footnote{Corresponding author: e-mail: {\sf a.malyshev@valbuena.fis.ucm.es}, 
	Phone: +34\,91\,394\,50\,12, 
	Fax:   +34\,91\,394\,45\,47}
}
\affiliation{
GISC, Departamento de F\'{\i}sica de Materiales, Universidad Complutense, Madrid, Spain
}

\begin{abstract}
It is argued that the (traditional) global level statistics which determines
localization and coherent transport properties of disordered systems at zero
temperature (e.g. the Anderson model)becomes inappropriate when it comes
to incoherent transport. We define local level statistics which proves to be
relevant for finite temperature incoherent transport and optics of
one-dimensional systems (e.g. molecular aggregates, conjugated
polymers, etc.).
\end{abstract}


\maketitle                   




\renewcommand{\leftmark}
{A. Malyshev: Local level statistics}

\section{Introduction} 

Level statistics has extensively been studied in the
connection with localization and transport properties of disordered systems
(see e. g. Ref.~\cite{Mirlin} and references therein). In particular,
nearest level spacing (NLS) distribution function $P(S)$ has a clear
physical meaning and its relationship with localization properties is easily
established: in the localized phase, energy levels are uncorrelated and
can be infinitesimally close in energy (the corresponding wave
functions can be localized far away from each other with a vanishing overlap),
which gives rise to the Poisson NLS distribution. On the other hand, in
the extended phase wave functions overlap well, which leads to strong level
repulsion and results in the Wigner-Dyson NLS statistics (time-reversal
symmetry is assumed hereafter). Various quantities associated with $P(S)$
were proposed as scaling variables to analyze metal-insulator transition
point, critical exponents, mobility edges, and other characteristics
relevant for coherent transport (see e. g. Ref.~\cite{Janssen} and
references therein). In the case of incoherent transport a quasi-particle can
hop between different localized states because of interactions with
environment (e.g. with a thermal bath as in the case of the phonon-assisted
hopping). The rate (or probability) of hopping between two states is then
proportional to the square of the matrix element of the interaction
calculated between wavefunctions of the states. Hopping probability between
states with vanishing overlap is therefore negligible (localized phase is
considered hereafter); scattering rate is relatively large only for the 
states that overlap well. The latter introduces correlation between pairs
of levels that are relevant for hopping.
No such correlation is accounted for in the traditional {\it global} level
statistics, we therefore define {\it local} level statistics that takes the
correlation into account and proves to be adequate for incoherent transport
in 1D disordered systems.

In the next section, we introduce a sample model of phonon-assisted
hopping, section 3 presents a definition of the {\it local} NLS
statistics in 1D, numerical results are discussed in section 4, while
section 5 concludes the paper.

\section{Diffusion model}

In this section we present an example model of quasi-particle diffusion
in a random uncorrelated potential \{$\varepsilon_n$\} mediated by weak 
coupling to a phonon system. The quasi-particle spectrum and wavefunctions 
\{$E_\nu, \Psi_\nu$\} are defined as:
\begin{equation}
\hat{H}\,\Psi_\nu = E_\nu\,\Psi_\nu,\quad
\hat{H}=
\sum_n^N \varepsilon_n\,|n\rangle\langle n| - 
t\sum_{n}^{N}\left(|n\rangle\langle n+1|+|n+1\rangle\langle n|\right)
\label{H}
\end{equation}
On-site energies $\varepsilon_n$ are randomly distributed within the box 
$[-W/2,W/2]$. All wave functions are localized at $W\neq 0$; incoherent
transport can then be modeled by means of the 
Pauli master equation for the populations $P_\nu$ of the quasi-particle
states:
\begin{equation}
    {\dot P}_\nu = R_\nu - \Gamma_\nu P_\nu + \sum_{\mu}
    (W_{\nu\mu}P_{\mu} - W_{\mu\nu}P_\nu) \ ,\qquad
    W_{\mu\nu} =   S(|\omega_{\mu\nu}|)\,G(\omega_{\mu\nu})
    \; I_{\nu\mu}\ .
\label{Pnu}
\end{equation}
where $R_\nu$ is the source term,
$\Gamma_\nu $ is the total decay rate of state $\nu$ (and describes the
drain), $W_{\nu\mu}$ is the scattering rate from the state
$|\mu\rangle$ to the state $|\nu\rangle$, which is due to the weak
coupling to a phonon system. The rates $W_{\nu\mu}$ are taken to be
proportional to the one-phonon spectral density $S(\omega)$ and 
the probability overlap integral $I_{\nu\mu} = \sum_{n} \Psi_{\mu n }^2
\Psi_{\nu n}^2$ (see e.g. Refs.~\cite{Malyshev03,Bednarz03}
and \cite{Shimizu98} for details),
$\omega_{\mu\nu} = E_\mu-E_\nu$, and $G(\omega)=n(\omega)$ if
$\omega>0$ while $G(\omega)=1+n(-\omega)$ if $\omega<0$, with
$n(\omega)=[\exp(\omega/T)-1]^{-1}$ being the mean thermal occupation
number of a phonon mode with energy $\omega$ (the Boltzmann constant is set to
unity while $T$ is the temperature). 
The model proved to be relevant for J-aggregates~\cite{Malyshev03,Bednarz03,Knoester02,Heijs05}.

\section{Local level statistics in 1D}

\begin{figure}[ht]
\includegraphics[clip,width=.53\textwidth]{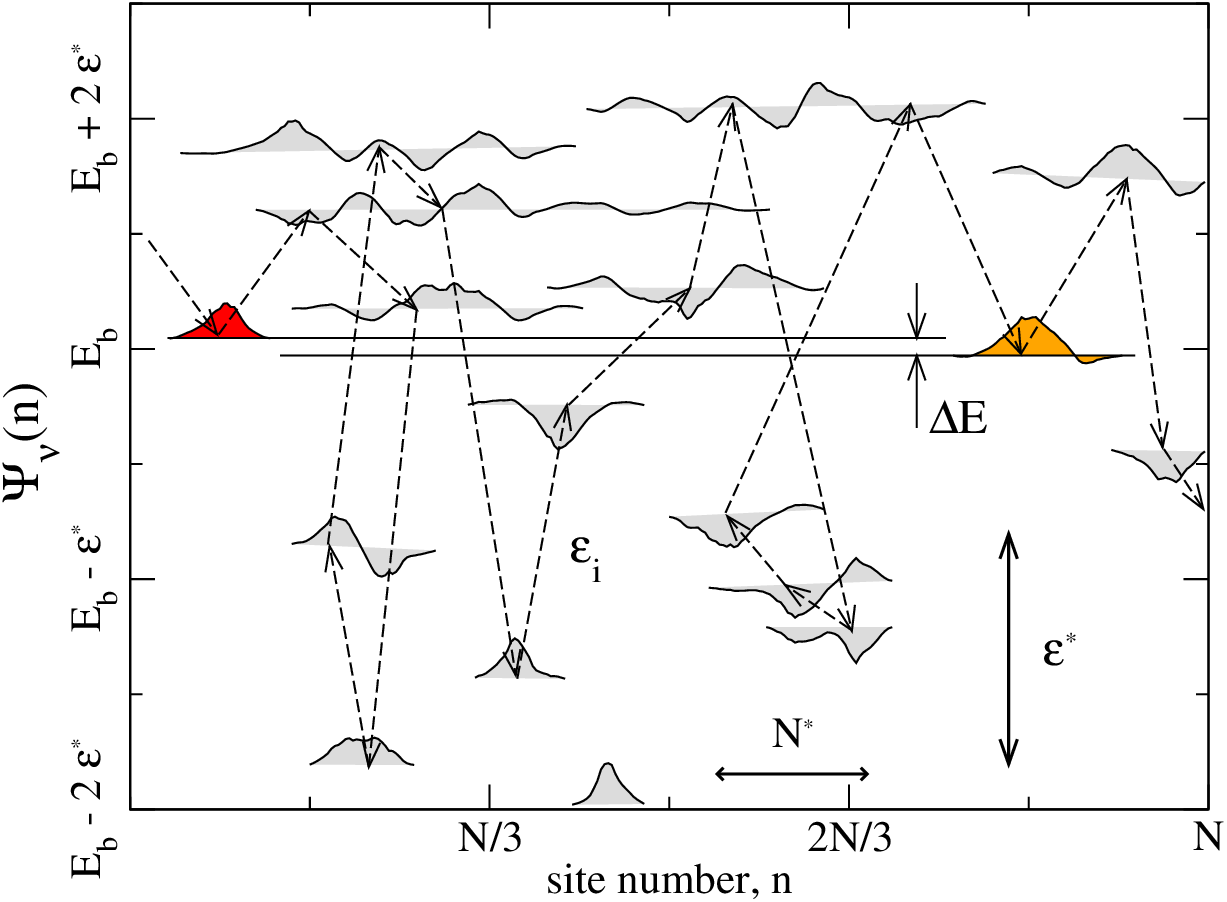}
\caption{
Wavefunctions and a diffusion trajectory: baselines of each wavefunction
represent the energy of the state, wavefunctions are in arbitrary units.
Wavefunctions forming the Lifshits tail of the density of states
($E_\nu \leq E_b = -2t$, $E_b$ is the bare band bottom [at $W=0$]) 
are localized at segments of typical size $N^*=N^*(W)$.
Dashed line --- a typical quasi-particle diffusion trajectory: because the hopping
probability is proportional to the the overlap integral $I_{\mu\nu}$ the
particle hops mostly over well-overlapping states. $\Delta E$ is the
smallest energy separation between two distant states, which is relevant for
global level statistics
($\lim_{N\to\infty}\Delta E=0$), $\epsilon_i$ is the energy spacing for 
the hop $i$ with the typical spacing $\epsilon^*$ such that 
$\lim_{N\to\infty}\epsilon^*=const$.
}
\label{fig:1}
\end{figure}

A typical realization of wavefunctions of the Hamiltonian (\ref{H})
for a chain of size $N$ and a possible diffusion trajectory is 
shown in Fig.~1. Note, that the overlap integral $I_{\mu\nu}$ that enters
the hopping rate $W_{\mu\nu}$ selects hops 
between states that overlap well (see also the figure caption for comments). 
The corresponding energy spacing at hop $i$ between such correlated
levels is $\epsilon_i$. To account for the correlation we propose the 
following procedure that is relevant for 1D case: (i) from the full sorted energy set \{$E_\nu$\} 
select energies within the window of interest
(e.g. $(-\infty,E_c]$ where the upper cut-off energy can reasonably be 
set as $E_c \sim E_b$ or $E_c \sim E_b + T$, $E_b$ being the bare band bottom 
[at $W=0$]), (ii) sort the selected energy set in the averaged
radius-vector for the state $\nu$, $\bar{x}_\nu = \sum_n|\Psi_{\nu n}|^2\,n$
(in 1D $\bar x$ is a number and such sorting is well defined);
let \{$\bar E_i:\,\bar{x}_{i+1}\geq\bar{x}_{i}$\} be the $\bar x$-sorted set, 
then (iii) construct a new energy sequence 
\{${\cal E}_n:\,{\cal E}_n=\sum_i^n s_i,\;
s_i=|\bar{E}_{i+1}-\bar{E}_i|$\}, i. e. new energies are 
cumulative sums of absolute spacings between $\bar x$-sorted energies.
The set \{${\cal E}_i$\} takes into account the above-mentioned correlations
because sequential states from the set \{${\cal E}_n$\} are always close 
in $r$-space, so that overlap integral between them is large. 
In the next section we
calculate the {\it local} NLS distribution functions $P_{E_c}(S)$ 
for $\bar x$-sorted energy sequences and discuss the results.

\section{Numerical results and discussion}

Fig.~2 demonstrates the {\it local} NLS statistics results calculated for $N=4096$ and
averaged over 1000 realizations of disorder. Left panel shows the dependence of the mean
NLS on the disorder magnitude $W$. In contrast to the global NLS statistics with vanishing 
mean spacing in the thermodynamic limit, the mean {\it local} NLS is independent of the 
system size. Calculated for $E_c=E_b$ it determines the characteristic 
energy scale for the states at the Lifshits
tail of the density of states, e.g., the diffusion activation energy at low
temperature (the latter was demonstrated for the case of the Frenkel exciton 
diffusion, see Ref.~\cite{Malyshev03}, where the statistics of the hidden energy
structure\cite{Malyshev01} was used; such statistics can be viewed as a
limiting (less general) case of the proposed {\it local} NLS statistics). 
The characteristic energy $\epsilon^*(W)$ of the {\it local} NLS for $E_c=E_b$ 
is in good agreement with the energy spacing in the hidden energy 
structure~\cite{Malyshev01}.

In order to analyze universal correlations between energy levels and 
compare NLS distribution functions to the standard ones (Poisson and Wigner-Dyson) it is 
necessary to unfold the raw energy sequence \{${\cal E}_n$\} 
(see Ref. \cite{Unfolding} for details). Right panel of Fig. 2 shows the distribution functions
$P_{E_c}(S)$ calculated from the unfolded energy sequence
for $W=0.3t$ and different cut-off energies $E_c$. 
Poisson and Wigner-Dyson distributions are shown for
comparison. When $E_c$ is well below the bare band edge $E_b=-2t$ ($E_c$ is deep in the Lifshits tail) 
the distribution tends to the Poisson distribution and demonstrates almost no level repulsion. 
In this case the states are localized far
away from each other and there is a vanishing overlap and correlation between
them. As the cut-off energy approaches the bare band bottom $E_b$ clear sign of
level repulsion appears in the {\it local} NLS distribution function. However, the
distribution does not tend to the Wigner-Dyson one upon increase of $E_c$, 
rather it tends to some limiting distribution. The latter is not surprising 
because although the $\bar x$-sorted sequence is correlated it is a sequence of localized states and no 
true Wigner-Dyson statistics can be expected. The reason for this is the following:
it turns up that there is a non-negligible amount of neighboring segments of the chain with 
very similar fluctuations of the random potential (averaged over the typical localization 
volume of the wavefunction, $N^*$);
energies of the states localized at such fluctuations are therefore very close to each other. 
Such pairs of fluctuations are similar to the system of two identical potential wells or a 
two-atom molecule; as the result, the "eigenstates" of such pairs of fluctuations remind bonded and anti-bonded 
states of two-atom molecule and can be very close in energy (subject to the wavefunction overlap). 
Positions of both states are also very close to each other, so they almost always
follow each other in 
the $\bar x$-sorted sequence making even the {\it local} NLS statistics less repulsive.
Quantitative contribution of such pairs of states to the NLS statistics will be analyzed elsewhere.

\begin{figure}[ht]
\includegraphics[clip,width=.49\textwidth]{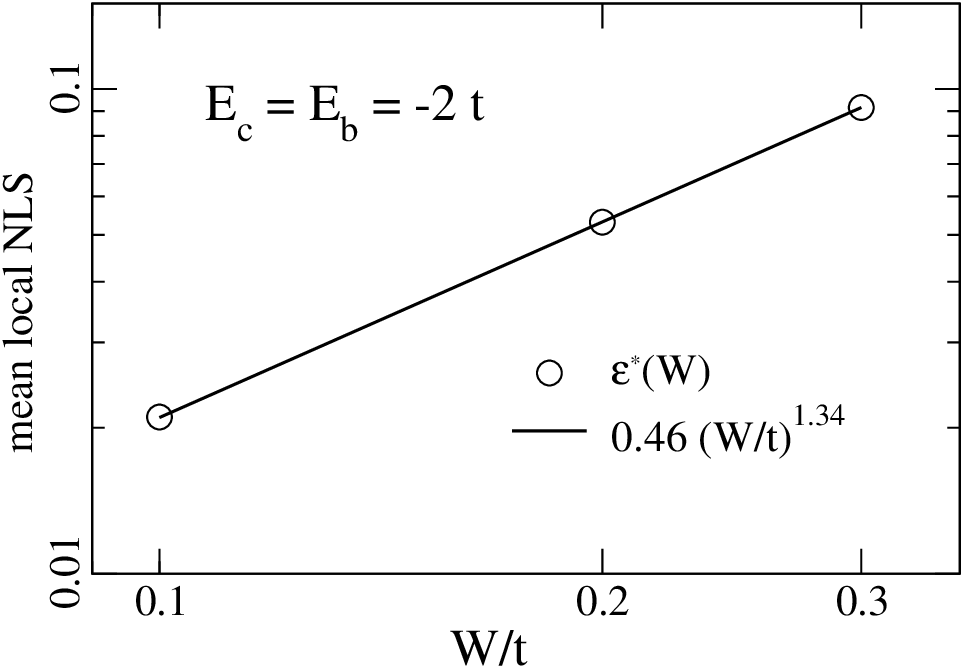}
\includegraphics[clip,width=.49\textwidth]{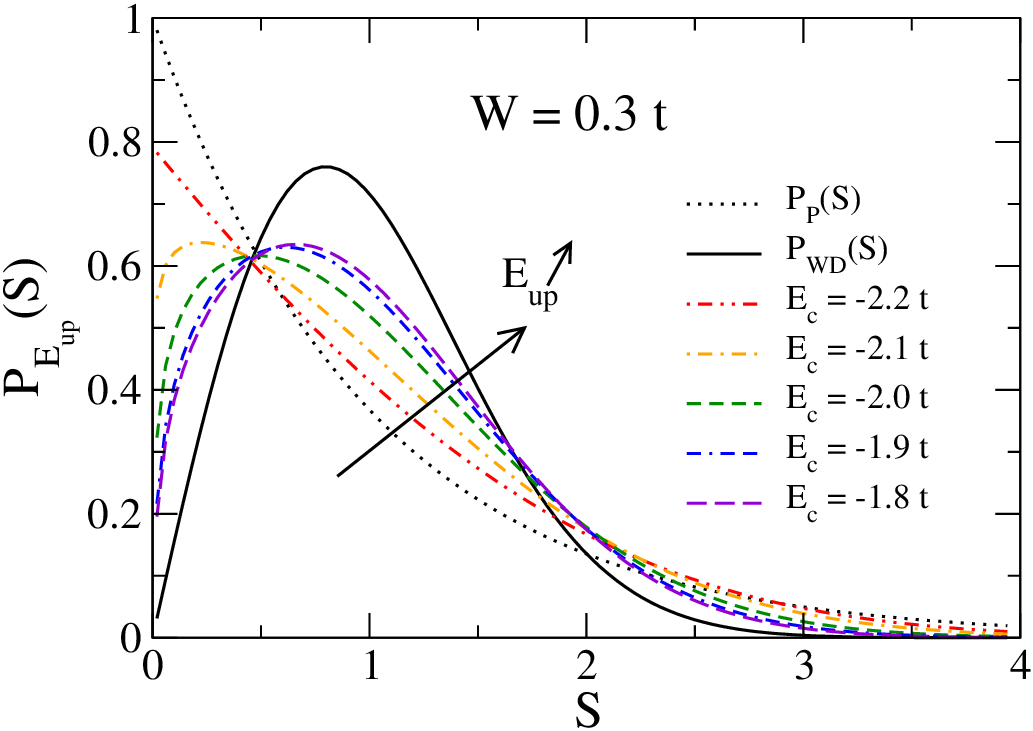}
\caption{
Left panel --- the dependence of the characteristic 
energy scale $\epsilon^*$ on the disorder magnitude $W$ for $E_c=E_b$.
Right panel --- {\it local} NLS distribution function $P_{E_c}(S)$ for the
unfolded $\bar x$-sorted sequence compared to the Poisson distribution
$P_P(S)$ (completely uncorrelated levels) and the Wigner-Dyson one $P_{WD}(S)$ 
(correlated levels). 
}
\label{fig:2}
\end{figure}

The dependence of the {\it local} NLS statistics on the cut-off energy
allows for definition of the diffusion mobility edge in a similar way 
the conventional mobility edge that separates extended and localized
phases is defined. Although no scaling variables or rigorous
argumentation is available in the present case (to the best of the author's knowledge),
by analogy, we formally define the diffusion mobility edge  
as the energy $E_\mu$ such that:
\begin{equation}
P_{E_{\mu}}(S_0)) - \frac{P_P(S_0)+P_{WD}(S_0)}{2} = 0\ ,
\end{equation}
where $P_P(S)$ and $P_{WD}(S)$ are Poisson and Wigner-Dyson distributions
respectively, and the spacing $S_0\sim 1$. In practice, we 
averaged $E_\mu$ over various $S_0:0.1\leq S_0\leq 3$. For all values of
disorder magnitude the diffusion mobility edge coincided with the
bare band bottom $E_b=-2t$. In order to interpret the later result it is
useful to compare the number of states below the mobility edge to the number
of the Lifshits tail states $N/N^*$ (see Fig.~1), 
it turns up that for all values of $W$
\begin{equation}
\int_{-\infty}^{E_\mu}\rho(E)\, dE \approx \frac{N}{N^*}\ , \qquad \rho(E)
=\left\langle \sum_\nu^N \delta(E-E_\nu)\right\rangle\ .
\end{equation}
The diffusion mobility edge has therefore a clear physical meaning: it separates
the spectral region with states that do not overlap with neighbors 
(hopping rates between such neighbors are vanishing)
from the energy region where wavefunctions begin to overlap well: higher (band) 
states cover one or more tail states and overlap well with them and between each 
other as well (see Fig.~1). In the latter case hopping rates and 
diffusion build up drastically.
The importance of two-step hops via higher states was discussed in detail in
Ref.~\cite{Malyshev03} in connection with the Frenkel exciton diffusion.

\section{Conclusions}

We have defined {\it local} NLS statistics for 1D disordered systems. The
{\it local} statistics grasps correlation between levels that are 
involved in hopping transport and is generally 
more repulsive than the global one. The {\it
local} NLS distribution in the localized phase changes from the Poisson one to 
the distribution that
demonstrates clear level repulsion, depending on the spectral region and the energy window.
Monitoring the change we determine the diffusion mobility edge that has
a clear physical meaning: it separates the states with vanishing probability 
of hopping between neighbors (slow incoherent transport) from the states with large inter-state 
hopping rate (fast transport regime). Using similar argumentation {\it local} level statistics can
be defined for higher-dimensional systems for which the common mobility edge can be defined. 
The mobility edge determined by the {\it local} statistics would however be different from the 
common one because the former is always the diffusion mobility edge. 
Analysis of higher-dimensional cases is beyond the scope of
this paper and will be discussed elsewhere.

\vspace{0.5in}

AM is grateful to J.\ Knoester and V.\ Malyshev 
for fruitful discussions. Support by INTAS through YSF 03-55-1545 and
Spanish Ministry of Education and Science through Ram\'{o}n y Cajal Program is
acknowledged. AM is on leave from Ioffe Institute, St. Petersburg, Russia.

\end{document}